\documentclass{ptapap}
\usepackage{amssymb}
\usepackage{sidecap}

\author{Przemys{\l}aw Walczak}[IAUWr]
\author{Jadwiga Daszy{\'n}ska-Daszkiewicz}[IAUWr]
\author{Alexey Pamyatnykh}[CAMK]
\author{Gerald Handler}[CAMK]
\author{Andrzej Pigulski}[IAUWr]
\author{BEST}[BEST]
\affil[BEST]{Bright Target Explorer (BRITE) Executive Science Team}

\affil[CAMK]{Nicolaus Copernicus Astronomical Center\\
  ul. Bartycka 18, 00--716 Warszawa, Poland}
\affil[IAUWr]{Instytut Astronomiczny Uniwersytet Wroc{\l}awski \\ ul. Kopernika 11, 51--622 Wroc{\l}aw, Poland}

\title{Interpretation of the BRITE oscillation spectra of the early B-type stars: $\nu$ Eri and $\alpha$ Lup}
\headtitle{$\nu$ Eri and $\alpha$ Lup}

\begin{document}

\maketitle

\begin{abstract}

$\nu$ Eridani is a well known multiperiodic $\beta$ Cephei pulsator which exhibits also the SPB (Slowly Pulsating B-type stars) type modes. Recent frequency analysis of the BRITE photometry of $\alpha$ Lupi showed that the star is also a hybrid $\beta$ Cep/SPB pulsator, in which both high and low frequencies were detected.

We construct complex seismic models in order to account for the observed frequency range, the values of the frequencies themselves and the non-adiabatic parameter $f$ for the dominant mode. Our studies suggest, that significant modifications of the opacity profile at the temperature range $\log{T}\in (5.0-5.5)$ are necessary to fulfill all these requirements.

\end{abstract}

\section{Introduction}

New observations of $\nu$ Eridani and $\alpha$ Lupi provided by BRITE-Constellation \citep{BRITE1} have revealed previously unknown frequencies. Most of the recently discovered oscillations cover low-frequency domain, corresponding to the high-order g-modes (SPB type). Some new high-frequency modes corresponding to the low-order g and p modes ($\beta$ Cep type) were also found.

$\nu$ Eri had already been observed during the extensive 2003-2005 world-wide observational campaigns \citep{Handler2004,J05,Aerts2004, deRidder2004}. The analysis of the campaign data revealed 14 pulsational frequencies. Two of them were low-frequency g-modes, 12 high-frequency p-modes. $\nu$ Eri became known as a hybrid $\beta$ Cep/SPB pulsator. In the recent BRITE observations, \citet{Handler2016} derived 17 pulsational frequencies. Seven of them were low-frequency g-modes. One frequency, 0.43 d$^{-1}$, discovered during the 2003-2005 campaigns was also found in the new data, while the other, 0.61 d$^{-1}$, was missing in the BRITE observations. Therefore, six of the seven new high-order g-modes were previously unknown. In the p-mode domain, two frequencies, 6.73 d$^{-1}$ and 6.22 d$^{-1}$, detected in the previous observations were not present in the BRITE data. No new p-modes were found.
The undetected modes can be explained by the relatively short e-folding time for the amplitude growth, that is of the order of a few dozens of years in the standard models.

The radial mode of $\alpha$ Lupi, 3.85 d$^{-1}$, has been known for many years, \citep[e.g.][]{Rodgers1962,Lampens1982,Mathias1994}. Our preliminary analysis of the BRITE observations of $\alpha$ Lup revealed 3 more p-modes. In addition, there were found many low frequencies in the range 0.27 - 0.7 d$^{-1}$. This means, that $\alpha$ Lup is one more hybrid $\beta$ Cep/SPB pulsating star.

Our aim was to calculate models that would reproduce observed features of the stars.
We were especially interested in explaining the instability in wide observed frequency ranges. It turned out, that the standard models cannot account simultaneously the instability of the low-frequency g-modes and high-frequency p-modes. Some modification of the opacity profile is needed to increase the driving effect. This results are in alignment with recent experiments \citep{Bailey_e2015}, where higher than predicted opacities of iron were measured.

In Sect.\,\ref{s2}, we summarized the basic information about the stars. In Sect.\,\ref{s3}, we described the standard and modified pulsational models. Conclusions ends the paper.

\section{Target stars}
\label{s2}

Both stars, $\nu$ Eri and $\alpha$ Lup are massive stars of the B2 spectral type. The effective temperature of $\nu$ Eri, $\log{T_{\rm{eff}}}=4.345\pm0.014$, was derived from \citet{JDD2005}. The luminosity, $\log{L/L_{\odot}}=3.886\pm0.044$, was determined with the parallax $\pi=4.83\pm0.19$ \citep{2007A&A...474..653V}.

In case of $\alpha$ Lup, the effective temperature $\log{T_{\rm{eff}}}=4.364\pm0.028$ was adopted from \citet{2009A&A...501..297Z}. The value of the parallax $\pi=7.02\pm0.17$ \citep{2007A&A...474..653V} corresponds to luminosity $\log{L/L_{\odot}}=4.180\pm0.064$.

\begin{SCfigure}
\centering
\includegraphics[width=0.5 \textwidth]{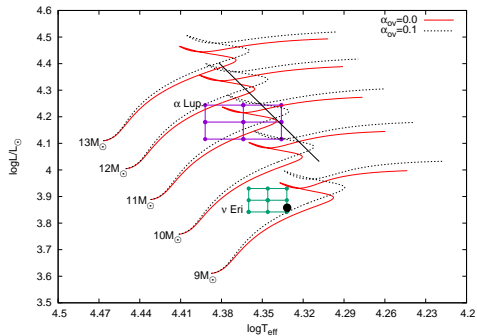}
\caption{The observational error boxes of $\nu$ Eri and $\alpha$ Lup in the  Hertzsprung-Russell diagram. The evolutionary tracks were computed for metallicity $Z = 0.015$ and two values of the overshooting parameter $\alpha_{\rm{ov}}=0$ and 0.1. A straight solid line indicates models that fit the frequency of the radial mode of $\alpha$ Lup. A model marked as a big dot near the $\nu$ Eri error box was described in the text.}

\label{fig:HR}
\end{SCfigure}

Both stars are shown in the Hertzsprung-Russell (HR) diagram in Fig.\,\ref{fig:HR}. The theoretical evolutionary tracks were calculated with the metallicity parameter $Z=0.015$, initial hydrogen abundance $X=0.7$ and two values of the overshooting parameter $\alpha_{\rm{ov}}=0$, 0.1. We assumed the solar chemical composition by \citet{AGSS09} and the new opacities from Los Alamos National Laboratory, OPLIB \citep{OPLIB2,OPLIB1}.

We can see, that the mass of $\nu$ Eri and $\alpha$ Lup is about $9.5M_{\odot}$ and $11.5M_{\odot}$, respectively. While former star seems to be well on the main sequence (MS), the evolutionary status of the latter is less established. Judging  by the position of the error box in the HR diagram, we may say that the star can either be on MS, undergo the contraction phase or have a hydrogen-burning shell around the helium core. To distinguish between these possibilities, we calculated the expected frequency change of the radial mode and compared it with an observed value. The theoretical frequency change rates are as follows: $\sim-5\times10^{-5}$ cycles per century on the MS, $\sim1\times10^{-3}$ cycles per century during the contraction phase and $\sim-1\times 10^{-2}$ cycles per century during hydrogen shell burning. The observed value of the radial mode of $\alpha$ Lup changed from $3.84842(4)$ d$^{-1}$ \citep{Lampens1982} to 3.84843(5) d$^{-1}$ in 2016 (this paper). This gives about $10^{-5}$ cycles per century. This value is most consistent with the MS hypothesis and in a further analysis we assumed, that $\alpha$ Lup is on MS.

\section{Pulsational Models}
\label{s3}

Mode identification of $\alpha$ Lup is rather uncertain. The only one firmly established mode degree correspond to the frequency $\nu_1=3.85$ d$^{-1}$, which is a radial mode ($\ell=0$).
Models of $\alpha$ Lup, that fit the frequency of its radial mode, calculated with $\alpha_{\rm{ov}}=0.1$, are marked in Fig.\,\ref{fig:HR} with a straight solid line. Assuming no overshooting and $Z=0.015$, we could not find a model simultaneously fitting the radial mode and lying inside of the error box.

\begin{figure}[h]
\includegraphics[width=6cm]{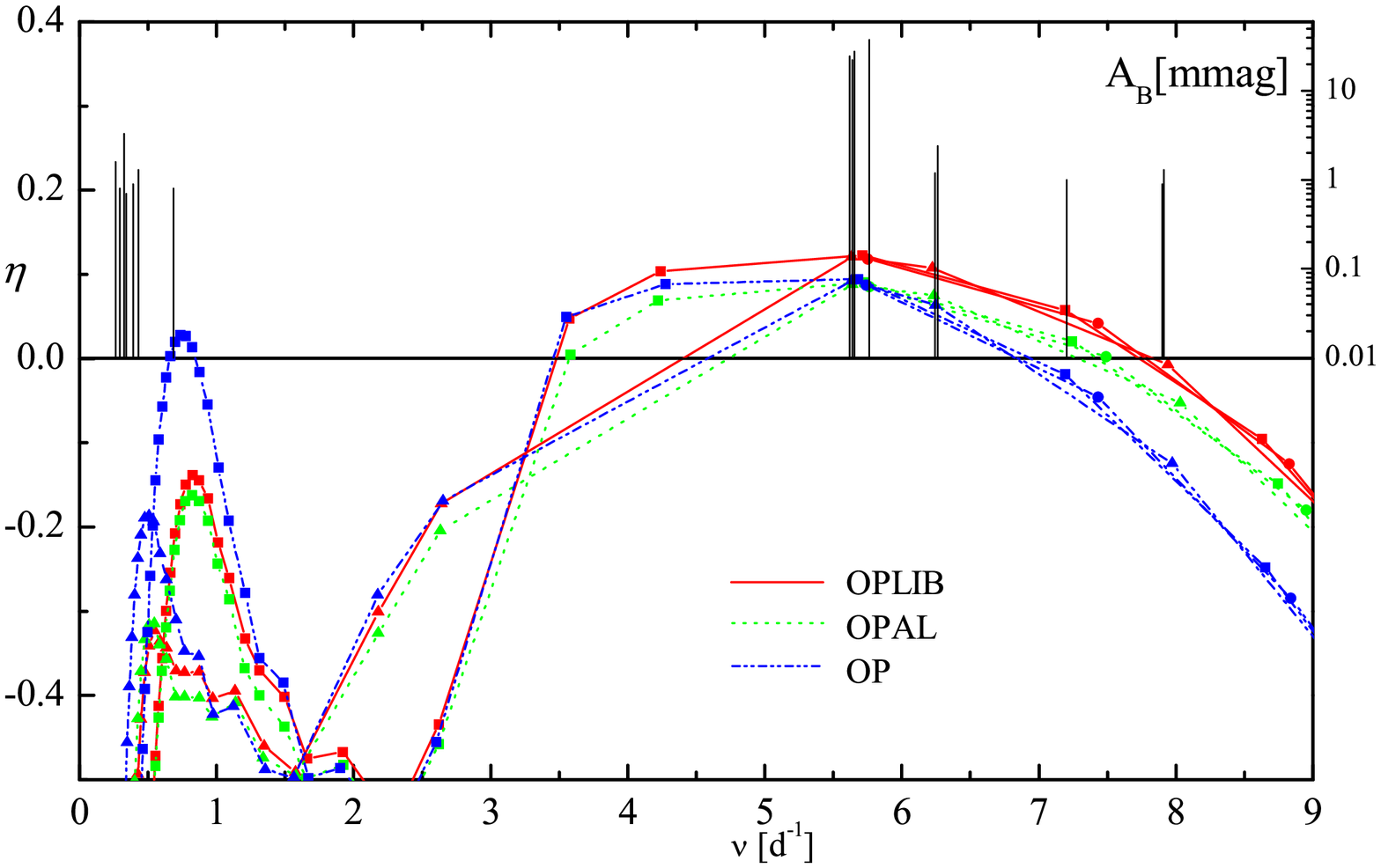}
\includegraphics[width=6cm]{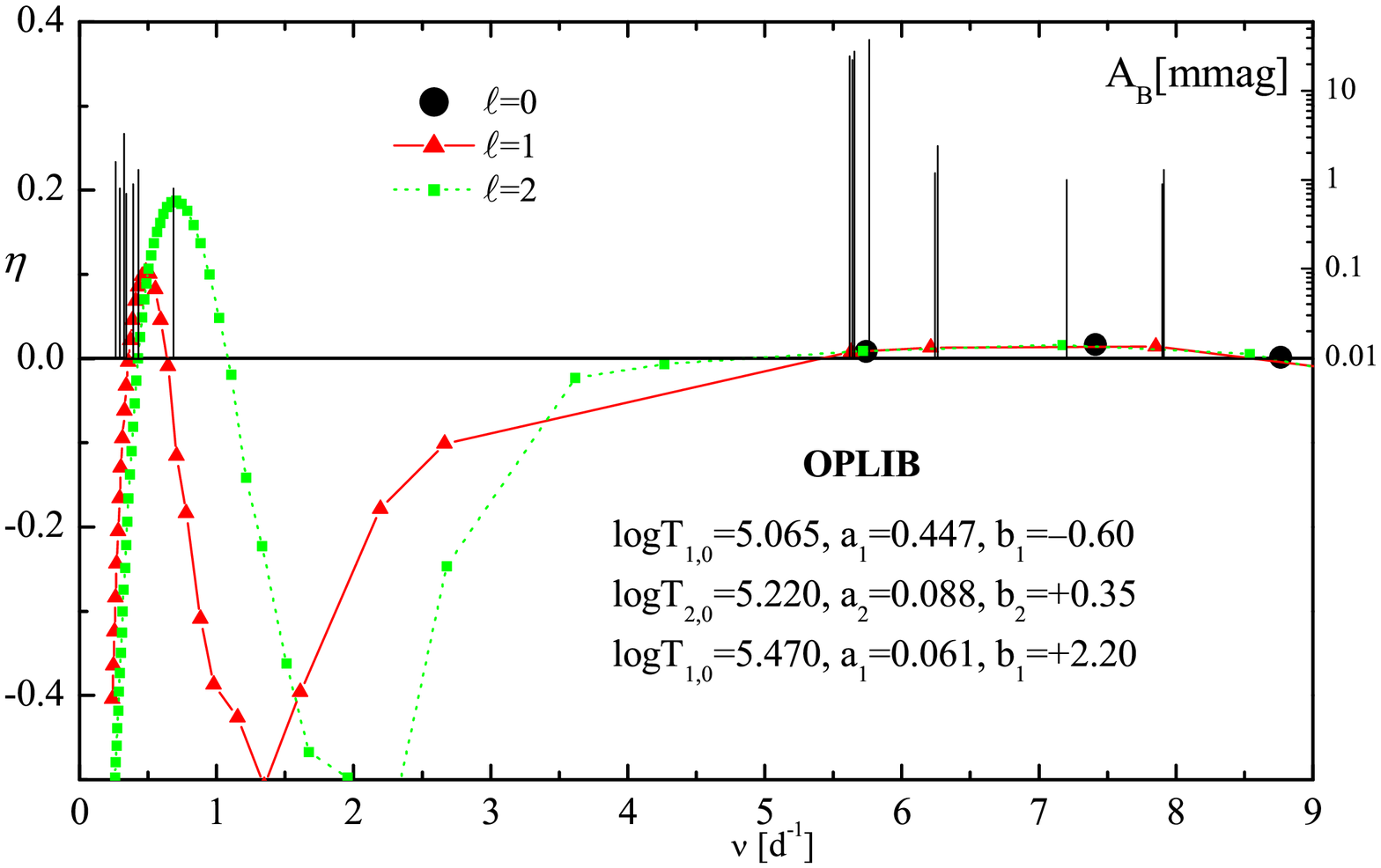}
\caption{Instability parameter, $\eta$, for mode degrees $\ell=0-2$ as a function of the frequency. The vertical lines indicate the observed frequencies of $\nu$ Eri. Their height corresponds to light variation amplitude in the BRITE B (blue) filter. In the left panel we show the standard models calculated with three different opacity tables: OPLIB, OPAL and OP. In the right panel a model with modified opacity profile  is shown. The modification coefficients are given in a legend (see text).}
\label{fig:nuEri}
\end{figure}

In the case of  $\nu$ Eri, we could use more frequencies in our modelling. We calculated models of $\nu$ Eri, that fit three frequencies, the radial mode (5.76 d$^{-1}$), and two centroids of the dipole modes (5.63 d$^{-1}$ and 6.24 d$^{-1}$). Such models calculated with three different standard opacity tables are shown in Fig.\,\ref{fig:nuEri} where we plotted the instability parameter, $\eta$, for mode degrees $\ell=0-2$. If $\eta>0$, the mode is unstable. The vertical lines indicate the BRITE oscillation spectrum. We used the OPLIB, OPAL \citep{OPAL} and OP \citep{OP} opacity tables. The models have metallicity $Z=0.015$, mass $M=9.5M_{\odot}$ and the overshooting parameter from 0.07 up to 0.09 ($\alpha_{\rm{ov}}$ was chosen to fit the frequencies). As we can see, none of standard models can explain the instability in the whole range of observed frequencies. The OPLIB model has unstable modes in the high-frequency domain, but low-frequency g-modes are not excited. The OPAL model is similar to the OPLIB model with somewhat smaller instability parameter especially near the high frequencies $\sim 7-8$ d$^{-1}$. The OP model has the highest instability for low-frequency g-modes, but still, the value of the instability parametr is both too small and shifted towards higher frequencies. The OP model cannot excite the high frequency modes with $\nu\gtrsim6.5$ d$^{-1}$, either.

Changing stellar parameters in a reasonable range does not improve the situation. Therefore, we decided to change artificially the standard opacity profile according to the following formula:
\begin{equation}
  \kappa (T)=\kappa_0(T) \left[1+\sum_{i=1}^N b_i \cdot\exp\left( -\frac{(\log T-\log T_{0,i})^2}{a_i^2}\right) \right],
\end{equation}
where $\kappa_0(T)$ is the original opacity profile and $(a,~b,~T_0)$ are parameters of the Gaussian describing the width, height and the position of the maximum, respectively.

It is possible to increase the instability parameter in regions of interest by changing the parameters $a$, $b$ and $T_0$. In fact, there is a large number of modified models, which can reproduce the observed frequency range. To constrain the number of combinations of the parameters $a$, $b$ and $T_0$ and to make the opacity modification more plausible, we included  the nonadiabatic parameter $f$ in our fitting. The parameter $f$ gives the ratio of the bolometric flux perturbation to the radial displacement at the level of the photosphere \citep{JDD2003,JDD2005} and its value is very sensitive to the opacity. The empirical values of $f$ can be derived from multicolor photometry and radial velocity measurements. Since the $f$-parameter is a complex quantity, it can be represented by the amplitude $|f|$ and the phase lag, $\Psi=arg(f)-180^{\circ}$. The phase lag $\Psi$ describes the phase shift between the maximum of the flux and the minimum of the radius. For the radial mode of $\nu$ Eri, the empirical values are $|f|=8.82(31)$, $\Psi=-4.82^{\circ}(1.98)$.  The small negative value of $\Psi$ means, that the maximum of the flux occur slightly after the minimum of the radius.

In many modified models, the theoretical values of the $f$-parameter differ significantly from the empirical counterparts. Only models with quite a complicated opacity modification were able to fit the unstable frequency range, the values of the three frequencies and the values of the $f$-parameter. One of our best models consists of the OPLIB data modification and it is shown in the right panel of Fig.\,\ref{fig:nuEri}. The model is also marked as a big dot in the HR diagram in Fig.\,\ref{fig:HR}. We see, that the model lies just on the border of the $\nu$ Eri error box.

The theoretical values of the $f$-parameter are $|f|=8.65$, $\Psi=-2.23^{\circ}$. The amplitude of $f$ is fitted within $1\sigma$ error, whereas the phase lag is reproduced within $2\sigma$ error. The model has unstable modes in the whole observed range of high frequencies. The observed low-frequency modes are also in unstable region. The only exceptions are the lowest frequencies. They can be, however, rotationally shifted to the smaller values of $\nu$. The opacity profile was modified at the three depths, $\log{T}$. There was a significant increase of the opacity at $\log{T_{0,3}}=5.470$ by 220\%. The second change was a small increase of $\kappa$ at $\log{T_{0,2}}=5.220$ by 35\%. In the uppermost layers we had to decrease the opacity at $\log{T_{0,1}}=5.065$ by 60\%.

\begin{figure}[h]
\includegraphics[width=6cm]{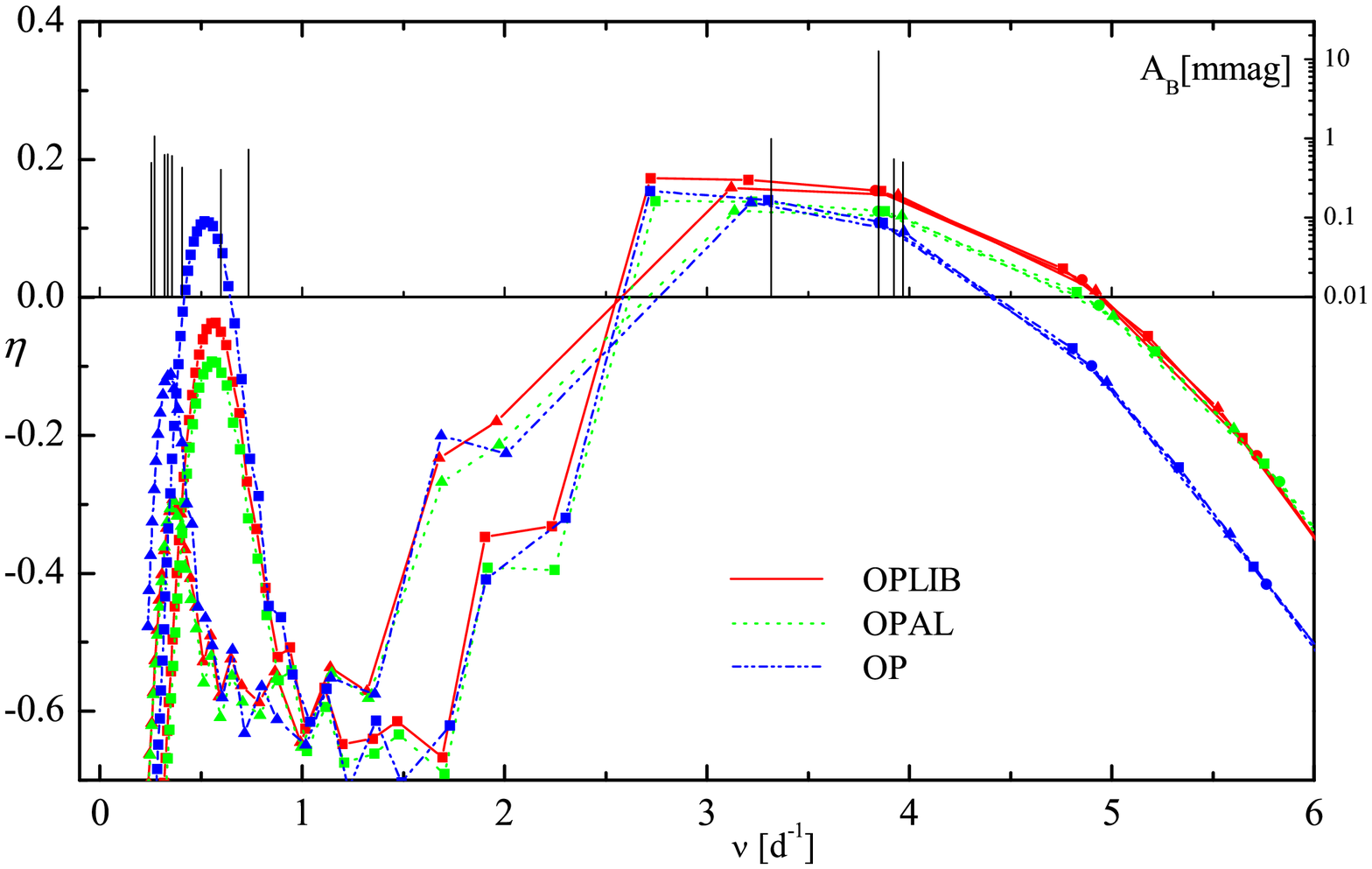}
\includegraphics[width=6cm]{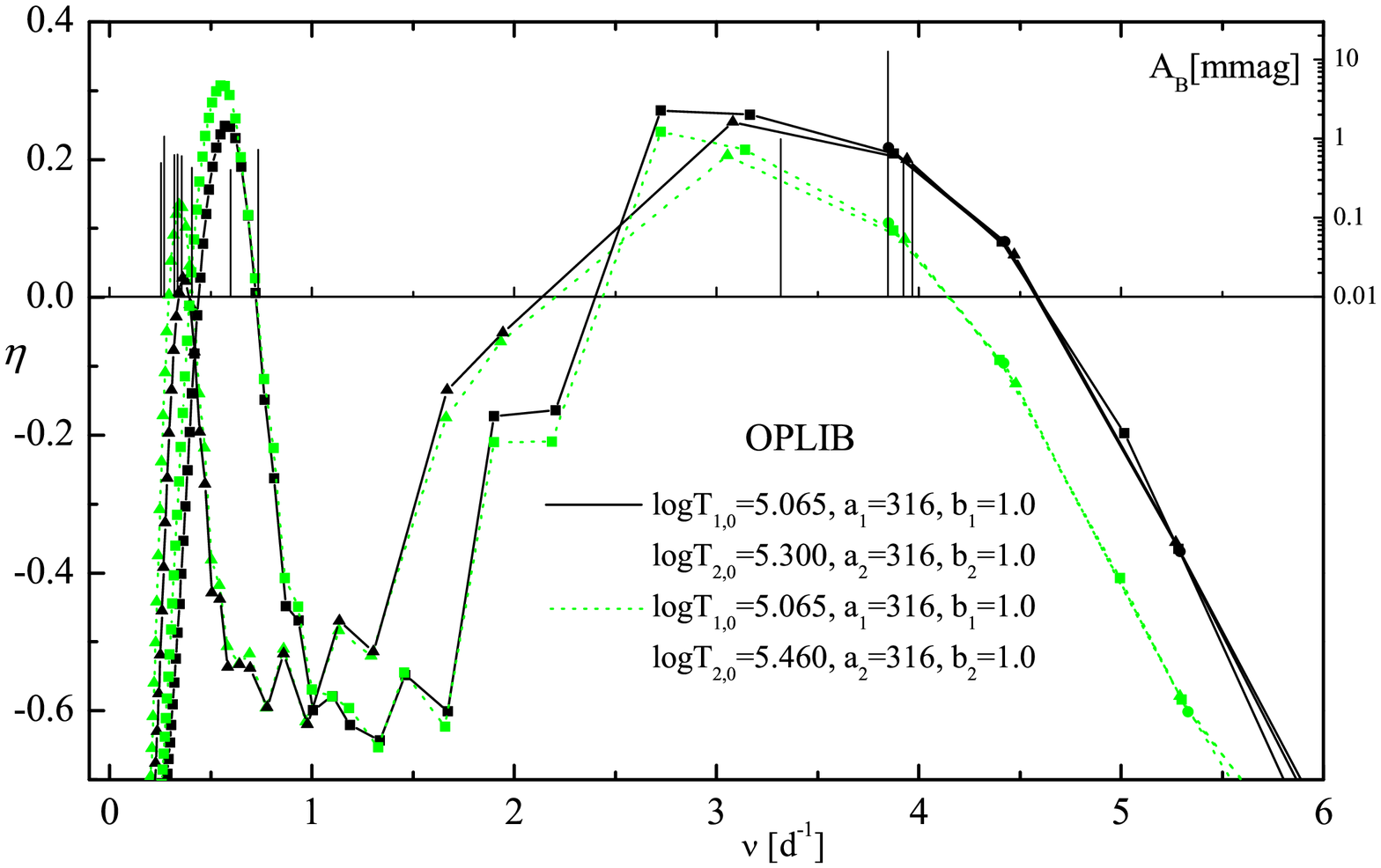}
\caption{The same as in Fig.\,\ref{fig:nuEri} but for the star $\alpha$ Lup.}
\label{fig:alphaLup}
\end{figure}

In the case of $\alpha$ Lup, we calculated models fitting the radial mode $\nu_1=3.85$ d$^{-1}$. Three examples of such models are shown in the left panel of Fig.\,\ref{fig:alphaLup}, which is the same as Fig.\,\ref{fig:nuEri} but for $\alpha$ Lup. All models have unstable modes in the high-frequency region. The standard OPLIB and OPAL models have stable high-order g-modes, whereas the OP model has unstable quadrupole modes ($\ell=2$).

In the next step we also computed pulsational models with modified opacities. Examples of such models are presented in the right panel of Fig.\,\ref{fig:alphaLup}, where we plotted the instability parameter for two different opacity modifications (the parameters are given in a legend).
Models with these modification seem to fit quite good the observed oscillation spectrum.
Unfortunately, we were not able to determine the empirical values of the $f$-parameter because of the lack of the radial velocities. We plan to do this in the near future. The perspectives are, however, promising because the theoretical values of the $f$-parameter for the radial mode of $\alpha$ Lup is very sensitive to the opacity.

\section{Conclusions}
\label{sc}

Our asteroseismic modelling of massive pulsators that exhibit both high-order g-modes and low-order g and p modes showed that standard models cannot account for the instability in the observed frequency ranges. As a possible solution, we considered modifications of the mean opacity profile. We used also the non-adiabatic $f$-parameter for the dominat radial mode to constrain the number of these modifications. The requirement of fitting simultaneously the observed range of frequencies, the values of individual frequencies and the values of the $f$-parameter significantly reduces the number of possible modifications.	

For $\nu$ Eri we found a model that nearly fulfills all the above  conditions. In that model a large increase of the opacity at $\log{T}=5.46$ is needed in order to get instability for low-frequency modes. Additionally, we had to reduce the opacities in the outer layers of the star to fit the empirical values of $f$ for the radial mode.  Such an opacity profile can result from either the intrinsic opacity of chemical elements and/or from inhomogeneous abundance pattern caused by different kinds of mixing.

For $\alpha$ Lup we calculated standard and modified models. The latter one seem to reproduce  pretty well the observed oscillation spectrum. Unfortunately, we did not derive the empirical values of $f$ so we could not control our modifications of $\kappa$. We are going to perform this analysis in the near future.

Our goal is to analyse more hybrid $\beta$ Cep/SPB pulsators (e.g. $\gamma$ Peg, 12 Lac) in a similar way in order to draw more general conclusions about preferable opacity modifications.

\acknowledgements{\newline This work was financially supported by the Polish NCN grant 2015/17/B/ST9/02082. PW's work was supported by the European Community's Seventh Framework Program (FP7/2007-2013) under grant agreement no. 269194. Calculations have been partly carried out using resources provided by Wroclaw Centre for Networking and Supercomputing (http://www.wcss.pl), grant No. 265. The paper is based on data collected by the BRITE Constellation satellite mission, designed, built, launched, operated and supported by the Austrian Research Promotion Agency (FFG), the University of Vienna, the Technical University of Graz, the Canadian Space Agency (CSA), the University of Toronto Institute for Aerospace Studies (UTIAS), the Foundation for Polish Science \& Technology (FNiTP MNiSW), and National Science Centre (NCN). The Polish contribution to the BRITE mission work is supported by the Polish NCN grant  2011/01/M/ST9/05914.}

\bibliographystyle{ptapap}
\bibliography{PWBiblio}

\begin{thebibliography}{19}
\providecommand{\natexlab}[1]{#1}
\providecommand{\url}[1]{\texttt{#1}}
\providecommand{\urlprefix}{URL }
\providecommand{\eprint}[2][]{\url{#2}}

\bibitem[{{Aerts} et~al.(2004)}]{Aerts2004}
{Aerts}, C., et~al., \emph{\mnras} \textbf{347}, 463 (2004)

\bibitem[{{Asplund} et~al.(2009){Asplund}, {Grevesse}, {Sauval}, \&
  {Scott}}]{AGSS09}
{Asplund}, M., {Grevesse}, N., {Sauval}, A.~J., {Scott}, P., \emph{\araa}
  \textbf{47}, 481 (2009), \eprint{0909.0948}

\bibitem[{{Bailey} et~al.(2015)}]{Bailey_e2015}
{Bailey}, J.~E., et~al., \emph{Nature} \textbf{517}, 56 (2015)

\bibitem[{{Colgan} et~al.(2015)}]{OPLIB2}
{Colgan}, J., et~al., \emph{High Energy Density Physics} \textbf{14}, 33 (2015)

\bibitem[{{Colgan} et~al.(2016)}]{OPLIB1}
{Colgan}, J., et~al., \emph{\apj} \textbf{817}, 116 (2016)

\bibitem[{{Daszy{\'n}ska-Daszkiewicz} et~al.(2003){Daszy{\'n}ska-Daszkiewicz},
  {Dziembowski}, \& {Pamyatnykh}}]{JDD2003}
{Daszy{\'n}ska-Daszkiewicz}, J., {Dziembowski}, W.~A., {Pamyatnykh}, A.~A.,
  \emph{\aap} \textbf{407}, 999 (2003)

\bibitem[{{Daszy{\'n}ska-Daszkiewicz} et~al.(2005){Daszy{\'n}ska-Daszkiewicz},
  {Dziembowski}, \& {Pamyatnykh}}]{JDD2005}
{Daszy{\'n}ska-Daszkiewicz}, J., {Dziembowski}, W.~A., {Pamyatnykh}, A.~A.,
  \emph{\aap} \textbf{441}, 641 (2005)

\bibitem[{{De Ridder} et~al.(2004)}]{deRidder2004}
{De Ridder}, J., et~al., \emph{\mnras} \textbf{351}, 324 (2004)

\bibitem[{{Handler} et~al.(2004)}]{Handler2004}
{Handler}, G., et~al., \emph{\mnras} \textbf{347}, 454 (2004)

\bibitem[{{Handler} et~al.(2016)}]{Handler2016}
{Handler}, G., et~al., \emph{\mnras} \textbf{subbmited} (2016)

\bibitem[{{Iglesias} \& {Rogers}(1996)}]{OPAL}
{Iglesias}, C.~A., {Rogers}, F.~J., \emph{\apj} \textbf{464}, 943 (1996)

\bibitem[{{Jerzykiewicz} et~al.(2005)}]{J05}
{Jerzykiewicz}, M., et~al., \emph{\mnras} \textbf{360}, 619 (2005)

\bibitem[{{Lampens} \& Goossens(1982)}]{Lampens1982}
{Lampens}, P., Goossens, M., \emph{\aap} \textbf{115}, 413 (1982)

\bibitem[{{Mathias} et~al.(1994)}]{Mathias1994}
{Mathias}, P., et~al., \emph{\aap} \textbf{283}, 813 (1994)

\bibitem[{{Rodgers} \& Bell(1962)}]{Rodgers1962}
{Rodgers}, A.~W., Bell, R.~A., \emph{Obs} \textbf{82}, 26 (1962)

\bibitem[{{Seaton}(2005)}]{OP}
{Seaton}, M.~J., \emph{\mnras} \textbf{362}, L1 (2005)

\bibitem[{{van Leeuwen}(2007)}]{2007A&A...474..653V}
{van Leeuwen}, F., \emph{\aap} \textbf{474}, 653 (2007), \eprint{0708.1752}

\bibitem[{{Weiss} et~al.(2014)}]{BRITE1}
{Weiss}, W.~W., et~al., \emph{PASP} \textbf{126}, 573 (2014)

\bibitem[{{Zorec} et~al.(2009)}]{2009A&A...501..297Z}
{Zorec}, J., et~al., \emph{\aap} \textbf{501}, 297 (2009), \eprint{0903.5134}

\end{thebibliography}

\end{document}